\documentclass[useAMS,usenatbib,usegraphicx]{mn2e}

\title[Single-pulse properties of AXP XTE J1810--197]{Simultaneous multi-frequency single-pulse properties of AXP XTE J1810--197}
\author[M.~Serylak et al.]{M.~Serylak$^{1,2}$\thanks{E-mail: serylak@astron.nl}, B.~W.~Stappers$^{3,1}$, P.~Weltevrede$^{4}$, M.~Kramer$^{3}$, A.~Jessner$^{5}$,\newauthor A.~G.~Lyne$^{3}$, C.~A.~Jordan$^3$, K.~Lazaridis$^{5}$ and J.~A.~Zensus$^{5}$\\
$^{1}$Netherlands Institute for Research in Astronomy, Postbus 2, 7990 AA Dwingeloo, The Netherlands\\
$^{2}$Astronomical Institute 'Anton Pannekoek', University of Amsterdam, Kruislaan 403, 1098 SJ Amsterdam, The Netherlands\\
$^{3}$Jodrell Bank Observatory, University of Manchester, Macclesfield, Cheshire, SK11~9DL, UK\\
$^{4}$Australia Telescope National Facility, CSIRO, P.O. Box 76, Epping, NSW 1710, Australia\\
$^{5}$Max-Planck-Institut f\"ur Radioastronomie, Auf dem H\"ugel 69, 53121, Bonn, Germany}
\begin{document}

\date{Accepted XXXX. Received XXXX; in original form XXXX}

\pagerange{\pageref{firstpage}--\pageref{lastpage}} \pubyear{2008}

\maketitle

\label{firstpage}

\begin{abstract}
We have used the 76-m Lovell, 94-m equivalent WSRT and 100-m Effelsberg radio telescopes to investigate the simultaneous single-pulse properties of the radio emitting magnetar AXP XTE J1810--197 at frequencies of 1.4, 4.8 and 8.35 GHz during May and July 2006. We study the magnetar's pulse-energy distributions which are found to be very peculiar as they are changing on time-scales of days and cannot be fit by a single statistical model. The magnetar exhibits strong spiky single giant-pulse-like subpulses, but they do not fit the definition of the giant pulse or giant micropulse phenomena. Measurements of the longitude-resolved modulation index reveal a high degree of intensity fluctuations on day-to-day time-scales and dramatic changes across pulse phase. We find the frequency evolution of the modulation index values differs significantly from what is observed in normal radio pulsars. We find that no regular drifting subpulse phenomenon is present at any of the observed frequencies at any observing epoch. However, we find a quasi-periodicity of the subpulses present in the majority of the observing sessions. A correlation analysis indicates a relationship between components from different frequencies. We discuss the results of our analysis in light of the emission properties of normal radio pulsars and a recently proposed model which takes radio emission from magnetars into consideration.

\end{abstract}


\section{Introduction}

Anomalous X-ray Pulsars (AXPs) and Soft Gamma-ray Repeaters (SGRs) form a class of slowly rotating neutron stars with rotational periods ranging from 2 to 12 seconds, called magnetars. Magnetars are characterised by properties such as: emission over a wide spectrum of wavelengths ($\gamma$-ray, X-ray, optical, near infrared and radio), rapid spin-down and a very high magnetic field (higher than $10^{14}$\,G, which is two orders of magnitude higher than for a radio pulsar of the same age and three orders of magnitude greater than for an average radio pulsar). These properties are very peculiar compared to the normal rotating neutron star behaviour and are best understood in the context of the magnetar model first discussed by \citet{dt92a}.

The AXP XTE J1810--197 was serendipitously discovered in 2003 as an X-ray pulsar by \citet{ims+04} in data taken with the {\it Rossi X-Ray Timing Explorer} {\it (RXTE)} while observing the known Soft Gamma-ray Repeater SGR 1806--20. A search in the archival {\it XTE} data showed that it produced an outburst around November 2002 with a persistent decline of its X-ray flux since then. The reported X-ray pulsar spin period was 5.54 s with a spin-down rate $\sim 10^{-11}$\,s\,s$^{-1}$ which, using standard methods, implied a magnetic field strength of $\sim 3\times10^{14}$\,G \citep{ims+04}. Due to this unusual long-term flux variability (outburst and exponential decline) it was classified as the first transient magnetar.

In 2004 a radio source with a strong flux density of $4.5\pm 0.5$\,mJy was found at the exact position of XTE J1810--197 by \citet{hgb+05} in data from the VLA MAGPIS survey at 1.4 GHz. Archival data were searched for any earlier detections but only upper limits could be estimated. Observations made by \citet{crh+06} at the Parkes radio-telescope and the Green Bank Telescope discovered the presence of the first pulsed radio emission from a magnetar. In a following series of papers, Camilo and collaborators investigate the magnetar's properties including: polarisation, flux density and pulse morphology variations in the radio regime (\citealt{ccr+07} \& \citealt{crj+07}) and broadband observations (radio, infrared to X-ray; \citealt{crp+07}). The results show large variations in pulse morphology and flux with an overall tendency for the flux to decrease during mid-2006 observations. In the latter paper the authors report the detection of single pulses at 88 GHz with the IRAM 30 m telescope and an infrared counterpart with variable flux and is uncorrelated with either X-ray or radio fluxes.

In their recent papers, \citet{ksj+07} and \citet{ljk+08} presented results from the first, full polarisation, simultaneous multi-frequency observations of XTE J1810--197. Although the magnetar shares some properties observed in ordinary radio pulsars, some distinct differences were also found. The most peculiar is the very flat and variable spectrum with an average spectral index $\alpha = 0.00 \pm -0.08$, which makes it the brightest neutron star emitting at frequencies above 20 GHz. Another difference is the variability of the pulse profile. Its extreme changes of shape on time-scales of less than a day, makes it quite unusual when compared to normal radio pulsars which typically need only several hundred individual pulses to obtain a stable pulse profile. Moreover the high degree of linear polarisation in the main pulse ($\sim 90\%$) and the position angle evolution with time provides evidence of a clear difference in the magnetar emission mechanism compared to normal radio pulsars.

Thanks to the initially extremely bright nature of XTE J1810--197 we are able to detect its single pulses, by analysis of which, we will try to understand the long-term pulse morphology and if it shows any correspondence with the single-pulse properties of normal radio pulsars. We report simultaneous multi-frequency single-pulse observations conducted at frequencies of 1.4, 4.9 and 8.35 GHz during observing sessions in May and July 2006. We present the results of our search for subpulse modulation of XTE J1810--197. We also investigate the pulse-energy distributions at all frequencies and epochs and the pulse-to-pulse correlation properties between the simultaneous single pulse observations. We then consider these properties in light of the emission characteristics of normal and giant pulses from ordinary pulsars.

\section{Observations}

The observations were made using the 94-m equivalent Westerbork Synthesis Radio Telescope (WSRT) in the Netherlands, the 76-m Lovell radio telescope at Jodrell Bank Observatory of the University of Manchester, UK and the 100-m radio telescope of the Max-Planck Institute for Radioastronomy (MPIfR) at Effelsberg, Germany, which are working in the European Pulsar Network (EPN) collaboration. The detailed discussion of the observing systems and calibration procedures is described by \citet{ksj+07}, while Table~\ref{table:summary of observing sessions} summarises the details of the observing sessions used in this paper.

\begin{table*}
\begin{minipage}{120mm}
\caption{Summary of observing sessions.}
\label{table:summary of observing sessions}
\begin{tabular}{cccccc@{\hspace{0.2cm}}c@{\hspace{0.2cm}}c@{\hspace{0.2cm}}c@{\hspace{0.2cm}}c}
\noalign{\smallskip}
\hline
\noalign{\smallskip}
\noalign{\smallskip}
Session & Date         & Telescope & Frequency & BW    & Total number & \multicolumn{4}{c}{Component} \\
        & MJD/dd.mm.yy &           & [GHz]     & [MHz] & of pulses    & \multicolumn{4}{c}{presence}  \\
\noalign{\smallskip}
\hline
\noalign{\smallskip}
1 & 53886/31.05.06 & Lovell     & 1.418 &   32 & 2101 & M1 & M2 & $-$ & $-$ \\
  &                & WSRT       & 4.901 &   80 & 2151 & M1 & M2 & M3 & $-$ \\
  &                & Effelsberg & 8.350 & 1000 &  972 & M1 & M2 & M3 & $-$ \\
\noalign{\smallskip}
2 & 53926/10.07.06 & Lovell     & 1.418 &   32 & 1947 & M1 & M2 & $-$ & $-$ \\
  &                & Effelsberg & 8.350 & 1000 & 2583 & M1 & M2 & M3 & IP \\
\noalign{\smallskip}
3 & 53933/17.07.06 & Lovell     & 1.418 &   32 & 2275 & M1 & M2 & M3 & IP \\
  &                & WSRT       & 4.896 &   80 & 3855 & M1 & M2 & $-$ & IP \\
  &                & Effelsberg & 8.350 & 1000 & 2454 & M1 & M2 & M3 & IP \\
\noalign{\smallskip}
\hline
\end{tabular}
\end{minipage}
\end{table*}

\section{Analysis and Results}

\begin{figure}
\includegraphics[scale=0.35,angle=270]{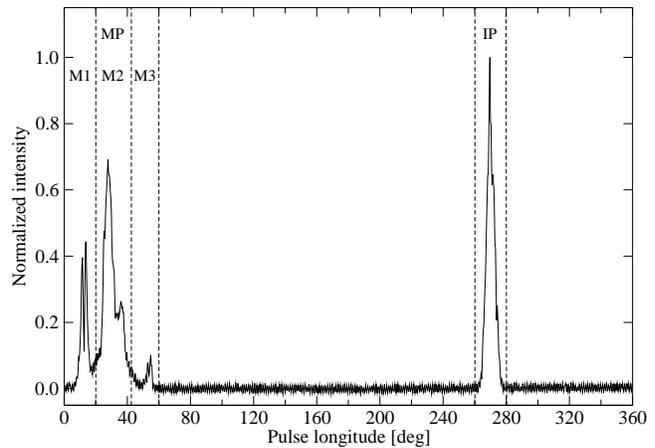}
\caption{An example of an average pulse profile of XTE J1810--197 from the observation made with the Effelsberg radio telescope during session 3. This plot shows the component naming convention used in this paper.}
\label{fig:average_profile}
\end{figure}

Observations of individual pulses from XTE J1810--197 allowed us to perform a variety of analysis techniques which we describe below. Before performing the analysis, all data sets have been corrected for any RFI present. In the case of the Lovell and WSRT observations re-binning was applied to increase the signal-to-noise ratio and to match the 5.4 ms time resolution of the Effelsberg data set. If more than one dataset was present per telescope per session, they were aligned in longitude and appended together to improve the statistics. This was done only for pulse-energy distributions, because any interruption (i. e. missing pulses) between consecutive data sets affects fluctuation and correlation analysis results. For the correlation analysis we aligned the data in phase and pulse number. Starting from session 2 we identify two emission regions in the average pulse profile, labelled as {\it main pulse} (MP), and {\it interpulse} (IP) according to the classification of \citet{ksj+07}. The MP region can be divided into at least three separate longitude regions containing separate components as marked in the panels in Fig.~\ref{fig:average_profile}. The average profile from the 1.4 GHz Lovell data in session 1 and 2 lacks the third component visible in the 4.9 GHz WSRT and 8.35 GHz Effelsberg data sets. The IP is not present in the first session and is only visible in the Effelsberg data in the second session. It is however visible in all three data sets in the third session. For the full list when the components are present see Table~\ref{table:summary of observing sessions}.

\subsection{The shape and stability of the pulse profile}

\begin{figure}
\includegraphics[scale=0.6,angle=270]{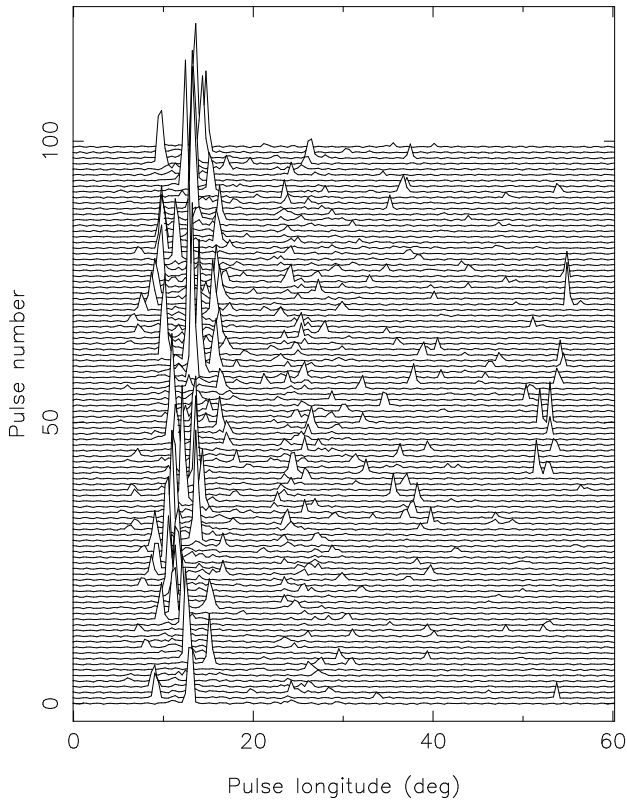}
\caption{An example of sequence of successive single pulses from the MP phase of XTE J1810--197 from the observation made at 8.35 GHz with the Effelsberg radio telescope during session 1.}
\label{fig:pulse_stack}
\end{figure}

Fig.~\ref{fig:pulse_stack} presents an example of a sequence of pulses from XTE J1810--197 observed at a frequency of 8.35 GHz during session 1. The plotted pulse-longitude range corresponds to the whole range of MP as can be seen in Fig.~\ref{fig:average_profile}. One can easily see that the subpulses are much narrower than the width of the average profile and appear at the longitude ranges corresponding to different components (see for instance the subpulses around pulse number 50). Also, it is worth noticing, that the subpulses associated with M1 are stronger than the subpulses from the remaining MP components, at any observed frequency and epochs, except the 8.35 GHz Effelsberg data set in session 3 where M2 becomes the strongest component in the MP longitude range (see Fig.~\ref{fig:average_profile}). The strong and spiky subpulses tend to be separated by $\sim 4$ deg ($\sim 61$ ms) throughout the data.

\subsection{Fluctuation analysis}

It was first shown by \citet{dc68} that for some pulsars, subpulses exhibit 'drifting' across the pulse window in an orderly fashion. This phenomenon forms 'drift bands' as can be seen in the 'pulse-stack' formed when one takes successive pulses and plots them on top of one other. The techniques used in this paper to investigate this phenomenon involve the Fourier transform and are known as the longitude-resolved fluctuation spectrum (LRFS; \citealt{bac70b}) and two-dimensional fluctuation spectrum (2DFS; \citealt{es02}). The procedure for the fluctuation analysis of our data sets is identical to those presented by \citet{wse07}, so we will present only a basic summary here.

\begin{figure*}
\includegraphics[scale=0.8]{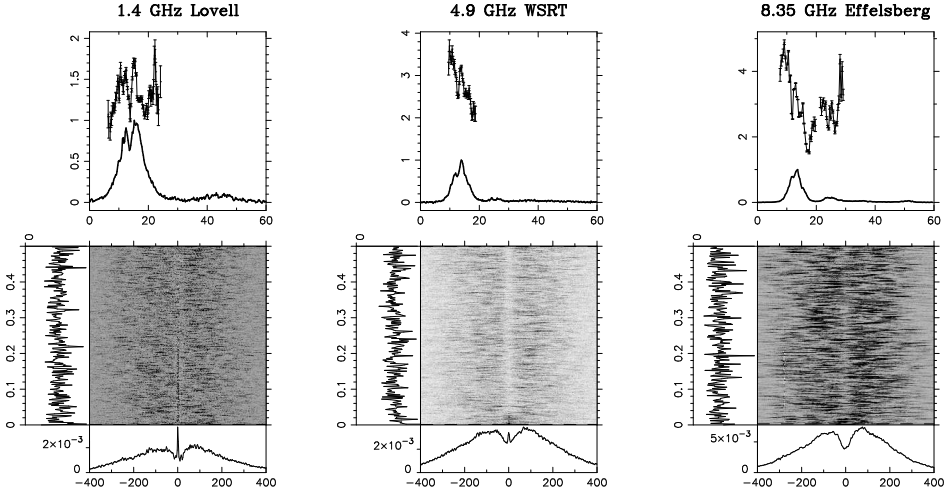}
\caption{Fluctuation analysis results shown for frequencies of 1.4 GHz ({\it left}), 4.9 GHz ({\it middle}) and 8.35 GHz ({\it right}) from session 1 (MJD 53886). The upper panels show the integrated pulse profile (solid line) with the peak amplitude normalised to 1, and the longitude-resolved modulation index, LRMI (solid line with error bars). The lower panels show the 2DFS where the ordinate of the resulting spectra are given in cycles per period (cpp) which corresponds to $P_{0}/P_{3}$ ($P_{0}$ is the pulsar period and $P_{3}$ denotes the vertical separation of possible drift bands) and the abscissa is also in cpp but now it corresponds to $P_{0}/P_{2}$ ($P_{2}$ denotes the horizontal separation of possible drift bands in longitude units). The greyscale intensity of 2DFS corresponds to the spectral power. The presence of a significant spectral feature with a value of $P_{3}$ and a positive or negative value of $P_{2}$, indicates that the subpulses appear with a preferred periodicity of $P_{3}$ pulsar periods and horizontal separation given in longitude units. The side panels correspond to horizontally (left panel) and vertically (bottom panel) integrated spectrum.}
\label{fig:all_telescope_2dfs_53886}
\end{figure*}

The first step in the fluctuation analysis is to form an average profile from a pulse-stack by vertically integrating each phase bin within the same longitude along consecutive pulses. Fig.~\ref{fig:all_telescope_2dfs_53886} shows the results from analysis at frequencies of 1.4, 4.9 and 8.35 GHz from session 1, where the average profile is drawn with the solid line in the top panel. The abscissa denotes the pulse phase in degrees. The top panel also presents the longitude-resolved modulation index (LRMI; solid line with error bars). The longitude-resolved modulation index is a basic method to estimate the presence of subpulse modulation. We only show LRMI values when the error in the LRMI is less than 0.5.

To investigate whether this modulation is systematic or lacks organisation, the 2DFS has to be calculated. This is done by dividing the pulse-stack into blocks of 512 pulses\footnote{Wherever the number of pulses was sufficient, otherwise shorter transforms was used.} and applying the Discrete Fourier Transform (DFT) along lines with different slopes in the pulse-stack. After averaging the spectra from each transformed block the final spectra is produced (Fig.~\ref{fig:all_telescope_2dfs_53886}, panels below average profiles). If the pulsar exhibits subpulse drifting this will be visible in its spectra as a region, so-called feature, of enhanced power in the greyscale. The vertical separation of possible drift bands is denoted by $P_{3}$ which is expressed in pulsar periods, $P_{0}$. The information about the horizontal separation of the possible drift bands, and thus whether the subpulses are drifting over a certain longitude range, is denoted by $P_{2}$ and is expressed in longitude units. A positive or negative $P_{2}$ value of a feature in the 2DFS denotes the separation of the subpulses, indicates that the subpulses appear later or earlier in successive pulses respectively.

We present the average profiles, LRMI and 2DFS from all telescopes from session 1 in Fig.~\ref{fig:all_telescope_2dfs_53886}. Results from our fluctuation analysis show no visible spectral features indicating that there is no regular drift patterns in any of our observations. However, one can see that there are spectral features in the 2DFS plots indicating subpulse modulation not associated with drifting subpulses, which will be discussed below. The LRMI in the 8.35 GHz Effelsberg data set shows a significant drop in the middle of M1, while towards its trailing edge the LRMI values increase. Inspecting the pulse stack indicates that this drop is caused by the frequent occurrence of many strong and narrow subpulses with similar intensities at that pulse phase. Similar behaviour appears in all the data sets in this session, but it is less visible in the 1.4 GHz Lovell and the 4.9 GHz WSRT data sets. In both the WSRT and the Effelsberg data sets the LRMI values are larger than the Lovell data set by a factor of two. The vertically integrated 2DFS (lower panels in Fig.~\ref{fig:all_telescope_2dfs_53886}) show similarity in shape across all the frequencies but change in intensity. There are two bumps and a strong peak in the middle of the vertically integrated 2DFS. We interpret the bumps as apparent subpulse separation, i.e. subpulses tend to be equally separated in the pulse profile throughout the observation. The peak is not related to the magnetar behaviour but indicates a baseline variations due to effects extrinsic to the source and instrumental effects due to the weather. This peak is also visible in the vertically integrated 2DFS from the WSRT data set, but it is weaker. In the Effelsberg data set the peak is no longer visible, while the bumps increase in intensity.

\begin{figure*}
\includegraphics[scale=0.8]{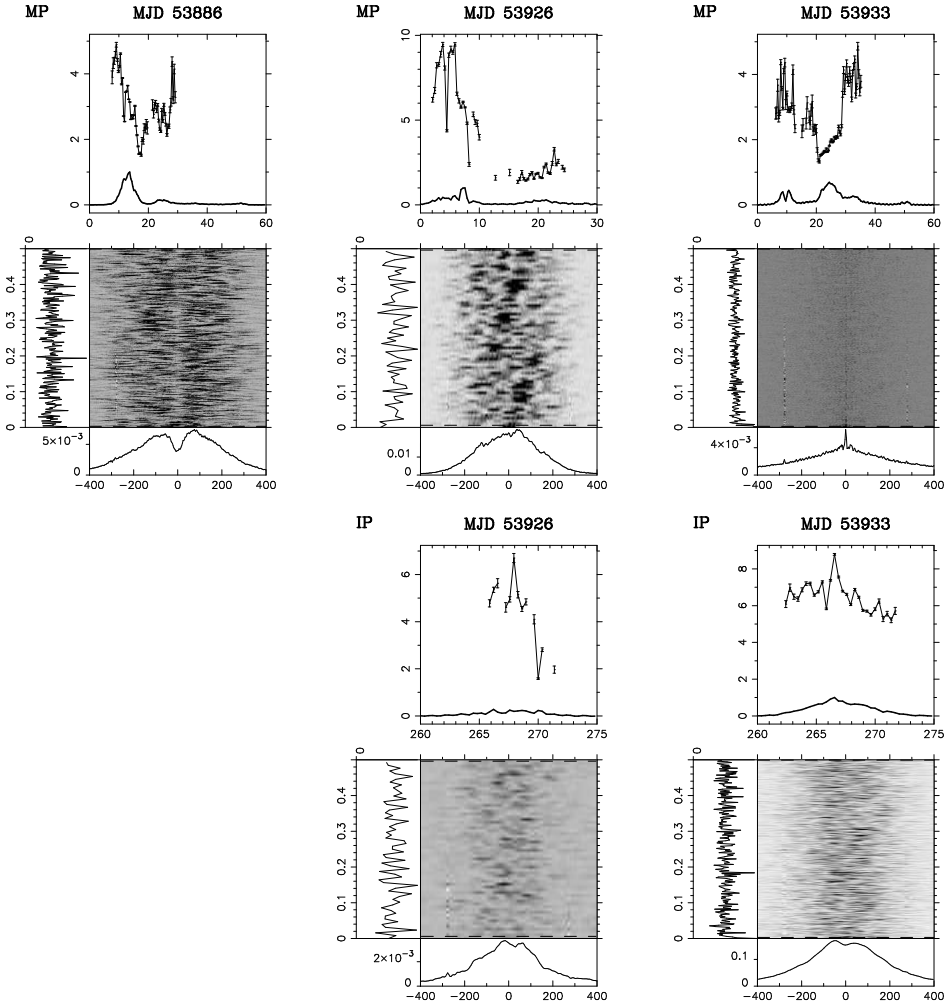}
\caption{Fluctuation analysis results shown from data obtained with the Effelsberg radio telescope at frequency of 8.35 GHz during all 3 sessions. The results from different longitude ranges covering the main pulse and the interpulse are shown in the upper and lower rows respectively. For MP from session 2 (MJD 53926) only first two components are shown. The notation is the same as in Fig.~\ref{fig:all_telescope_2dfs_53886}.}
\label{fig:effberg_2dfs_all_days}
\end{figure*}

In session 2, the 2DFS plots show no visible spectral features indicating there is no preferred periodicity. The average profiles indicate small changes from session 1 to session 2 indicating that the subpulse separation is now less regular. The 2DFS plots made from the 8.35 GHz Effelsberg data sets for both the MP and IP show a strange wave-like pattern (Fig.~\ref{fig:effberg_2dfs_all_days}, middle panel). Further analysis showed that this rapid pattern is due to interference and may be caused by the convolution of the interfering signal from the alternating stray magnetic fields of the fans used to cool the telescope receivers with the actual on-pulse magnetar signal. The pattern is stronger in the IP 2DFS plots due to the lower IP on-off-pulse signal ratio. This can easily be seen in the off-pulse region of the average profile of the MP (Fig.~\ref{fig:effberg_contour_all_days_mp}, lowest panel).

The LRMI values increase by a factor of about four in the 1.4 GHz Lovell data sets and a factor of two in the 8.35 GHz Effelsberg data sets for the two first components compared to those measured in session 1. Moreover, instead of the dip in the LRMI values shown in the previous session there is only a peak indicating high variability in the intensity of the observed pulses at the same pulse phase. The modulation indices for the Effelsberg IP tend to be lower towards the trailing edge of the IP profile. Due to the presence of the IP, there are two 2DFS plots, in Fig.~\ref{fig:effberg_2dfs_all_days}. The features visible in the vertically integrated 2DFS plots from session 1 do not appear in session 2.

In session 3 the 1.4 GHz Lovell data set (not shown) demonstrates a significantly large value of the LRMI at the trailing edge of the MP first component. This indicates infrequent emission of very strong subpulses. The LRMI for the IP (not shown) was not calculated because of the large error in the measurement in modulation indices due to a small number of bright pulses. Because of the low S/N ratio in the 4.9 GHz WSRT observations, there are only a few values of the LRMI calculated for the MP and none for the IP. The 8.35 GHz Effelsberg data set LRMI for the MP is two times lower than for the session 2 data set, hence is similar to the data set from session 1 (Fig.~\ref{fig:effberg_2dfs_all_days}, right panel). The M2 LRMI values become lower than those in M1 and M3. This indicates smaller intensity fluctuations of subpulses appearing in the middle component and larger variations on the side components throughout the observation. The LRMI for the IP rises until it peaks in the middle of the average IP profile, which means that there are large intensity variations at this longitude.

\begin{figure*}
\includegraphics[scale=0.6,angle=270]{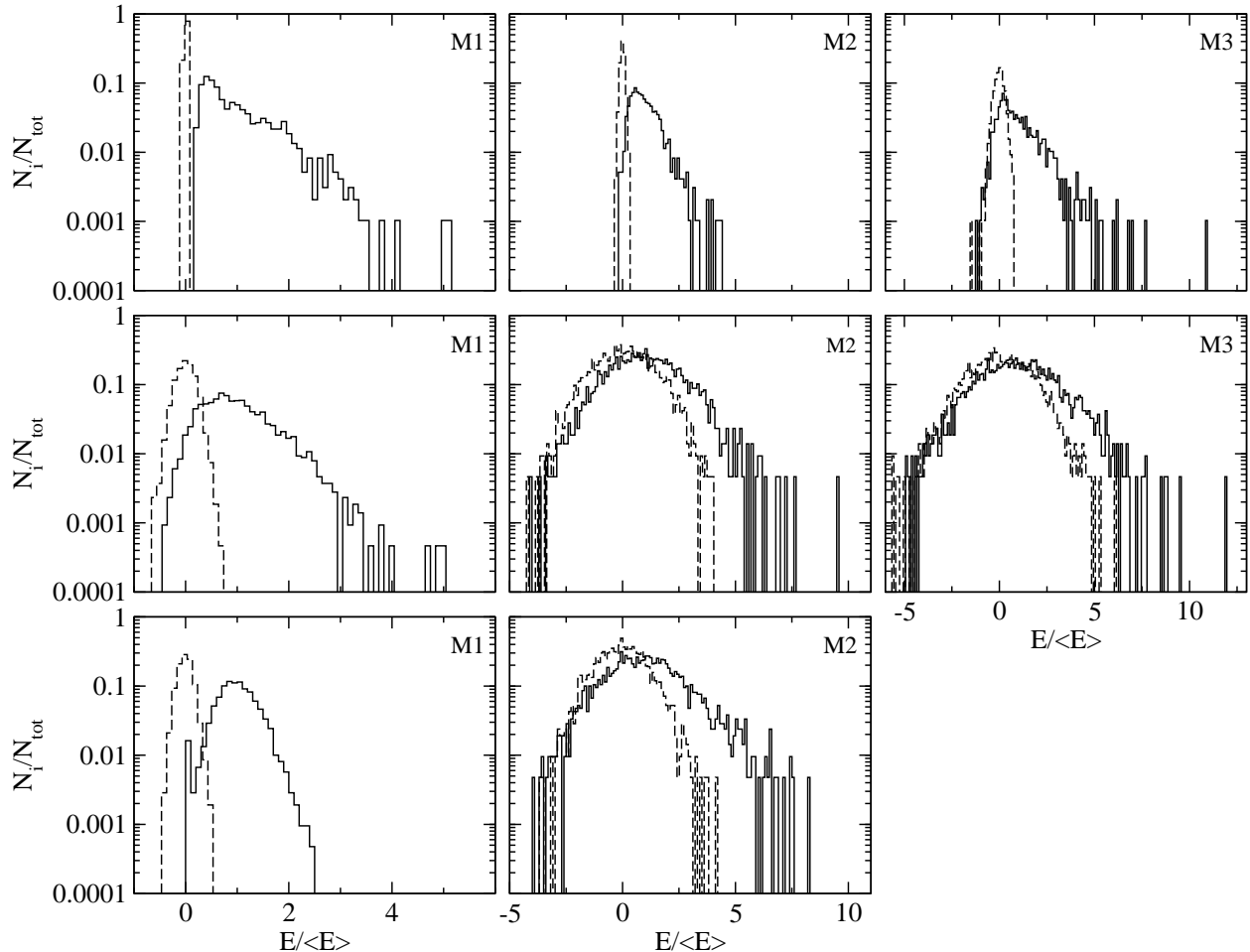}
\caption{Main pulse pulse-energy distributions from the 1.4 GHz Lovell ({\it bottom row}), 4.9 GHz WSRT ({\it middle row}) and 8.35 GHz Effelsberg ({\it top row}) observations from session 1. The components of the main pulse region are compared starting from first (M1; {\it left column}) through second (M2; {\it middle column}) ending with third component (M3; {\it right column}). The horizontal axis denote energies which are normalised and scaled to the average pulse energy of the component. The off-pulse energy distributions of the components are shown with dashed lines.}
\label{fig:all_telescope_penergy_53886}
\end{figure*}

\subsection{Pulse-energy distributions}

The results of the fluctuation analysis described in the previous section clearly give evidence of intensity fluctuations over short periods of time, even within the same observing session. To study the pulse-amplitude characteristics of XTE J1810--197 we have made pulse-energy distributions for all frequencies and epochs for both the MP and the IP. Then, for each of the components of the MP and for the IP we fit their pulse-energy distributions with two well known models. Because the emission of XTE J1810--197 shows strong and spiky giant-pulse-like subpulses, we used power-law statistics to model this behaviour as shown by \citet{lcu+95}, while for the 'normal' pulses the lognormal statistic was used as discussed by \citet{cjd01}:

\begin{equation}
P_{\mathrm{powerlaw}}(E) \propto E^{p},
\label{eq:powerlaw}
\end{equation}

\begin{equation}
P_{\mathrm{lognormal}} = \frac{<E>}{\sqrt{2\pi}\sigma E} \exp \left[- \left( ln \frac{E}{<E>} - \mu \right)^2 / (2 \sigma^{2})\right].
\label{eq:lognormal}
\end{equation}

Since the power-law distribution integral is infinite, we introduced the parameter $E_{min}$, a minimum energy of the pulses, which together with {\it p} is used for fitting the power-law. The lognormal model is fit with $\mu$ and $\sigma$. The average energy of the model distribution, $<E>$ was set to match that of the observed distribution. In our analysis, we also took into account the effect of the noise. Since the noise in some of our observations was not pure Gaussian, due to the RFI present (e.g. 8.35 GHz Effelsberg data sets) we convolved the observed noise with the model distributions. The derived parameters were optimised for the best fit by minimising the $\chi^{2}$ using the Amoeba algorithm \citep{pftv86} and are compared in Table~\ref{table:pulse energy fits}. The detailed description of the procedure of the fitting is well described by \citet{cai04} and \citet{wws+06}. We discuss specific cases below, but we note here that in general it was not possible to fit the distributions well with a single model.

Fig.~\ref{fig:all_telescope_penergy_53886} presents the main pulse energy distributions for all the telescopes from session 1. The lowest row shows pulse-energy distributions from 1.4 GHz with ascending frequency towards the top row. The components of the main pulse region are compared starting from M1 (left column) through M2 (middle column) ending with M3 (right column). For this session, only the M1 from the Lovell and WSRT data sets is best fit by lognormal pulse-energy distributions. The remaining components and the whole 8.35 GHz Effelsberg data set follows power-law-like statistics. The Effelsberg M1 is very strong and separates itself from the noise in the pulse-energy distribution (Fig.~\ref{fig:all_telescope_penergy_53886}; top left-most plot). We emphasise that the Effelsberg M1 pulse-energy distribution is clearly a power-law without weaker underlying emission. This is different from the case of normal radio pulsar emission, where the  of giant pulses result in power-law tail extending from the pulse-energy distribution.

\begin{figure*}
\includegraphics[scale=0.6,angle=270]{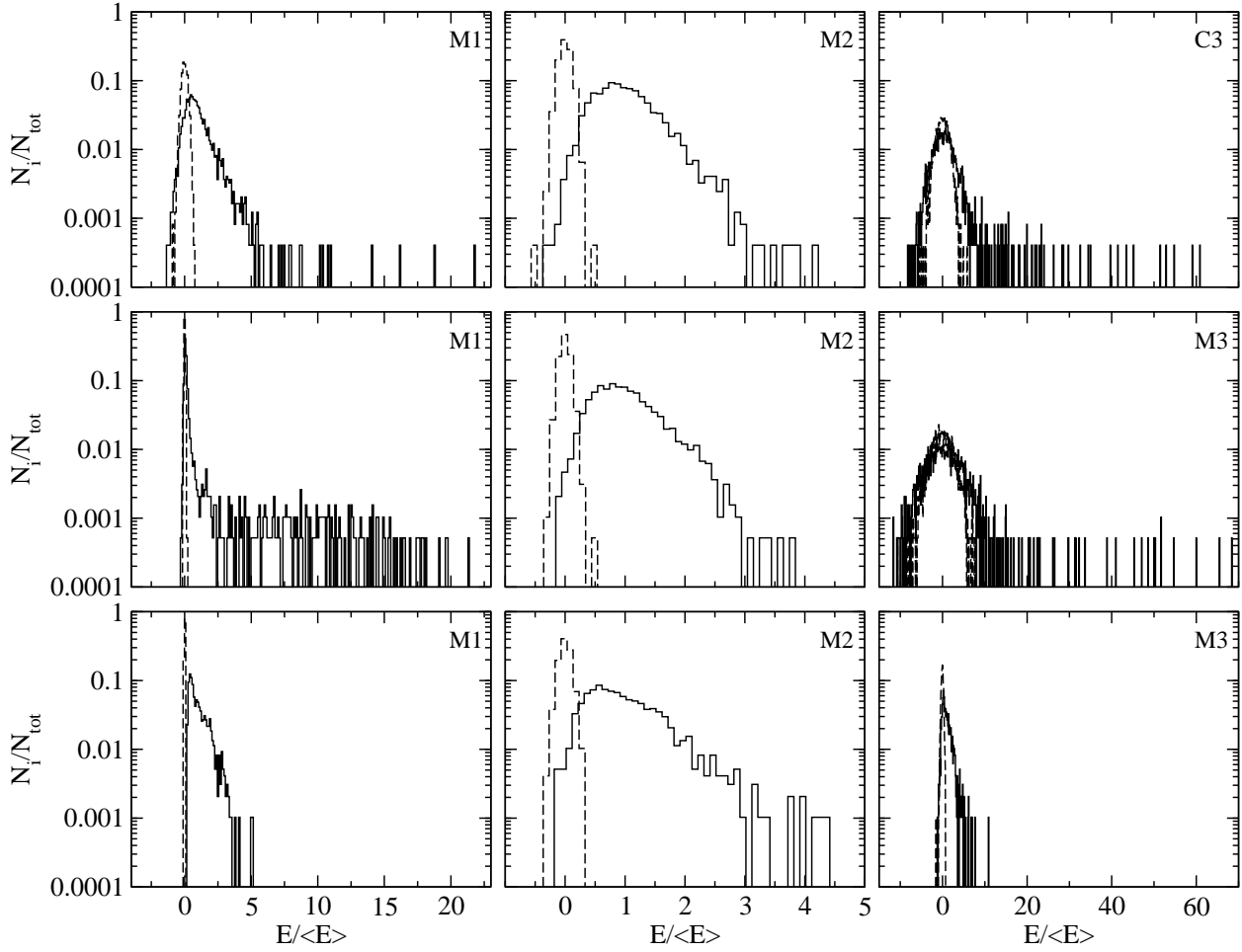}
\caption{Main pulse energy distributions from the 8.35 GHz Effelsberg radio telescope observations from session 1 ({\it bottom row}), session 2 ({\it middle row}) and session 3 ({\it top row}). The components of the main pulse region are compared starting from first (M1; {\it left column}) through second (M2; {\it middle column}) ending with the third component (M3; {\it right column}). The horizontal axis denote energies which are normalised and scaled to the average pulse energy of the component. The off-pulse energy distributions of the components are shown with dashed lines.}
\label{fig:effberg_penergy_all_days_mp}
\end{figure*}

\begin{figure*}
\includegraphics[scale=0.6,angle=270]{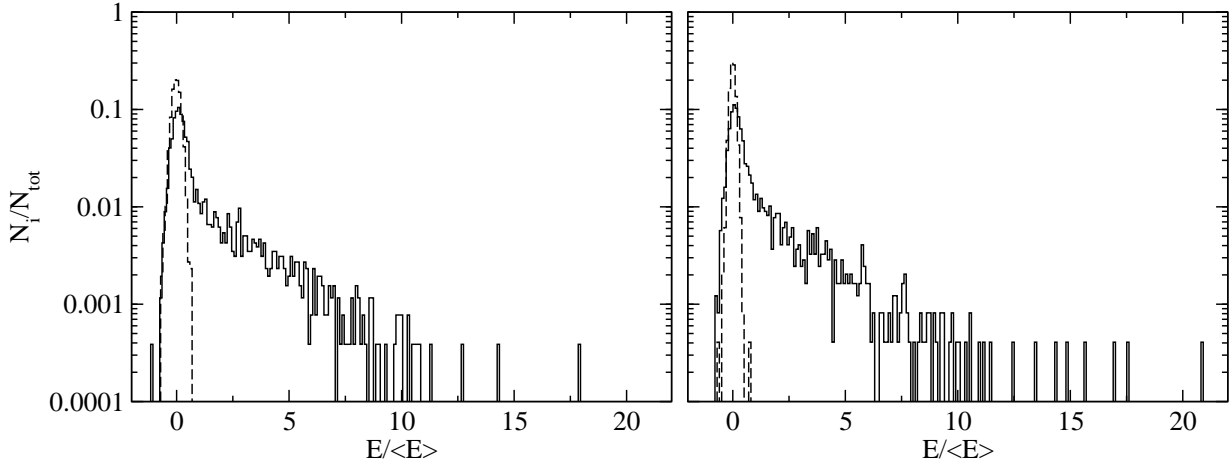}
\caption{The interpulse energy distributions from the 8.35 GHz Effelsberg radio telescope observations from session 2 ({\it left}) and session 3 ({\it right}). The off-pulse energy distributions are dashed lines. The horizontal axis denote energies which are normalised and scaled to the average pulse energy of the component. The off-pulse energy distributions of the components are dashed lines.}
\label{fig:effberg_penergy_all_days_ip}
\end{figure*}

In session 2 the 1.4 GHz Lovell pulse-energy distribution is best fit by lognormal behaviour. However, it also shows a long tail in the M1 pulse-energy distribution. M1 and M3 in the 8.35 GHz Effelsberg data set show similar behaviour (Fig.~\ref{fig:effberg_penergy_all_days_mp}). This introduces some complexity to the pulse-energy distribution fitting. Those pulse-energy components with long tails, cannot be fit with a single fit of either power-law or lognormal distribution. To try to get better fits, the distributions were split into two halves, the weak and strong pulses in terms of intensity, and then both halves of the distribution were fit with either lognormal or power-law distributions, like the giant pulse tails of normal radio pulsar distributions are usually fit \citep{lcu+95}. The Effelsberg IP (Fig.~\ref{fig:effberg_penergy_all_days_ip}; left plot) shows a peak and a flat tail in its pulse-energy distribution, but it is possible to fit its pulse-energy distribution with a lognormal pulse-energy distribution.

\begin{table*}
\begin{minipage}{115mm}
\caption{The parameters of the best fits of the model distributions. Besides power-law, {\it p} and lognormal, $\mu$, $\sigma$ distribution parameters, also the $\chi^{2}$, the number of degrees of freedom, $N_{d.o.f}$ and significance probability, $P(\chi^{2})$ are shown.}
\label{table:pulse energy fits}
\begin{tabular}{cccccccl}
\noalign{\smallskip}
\hline
\noalign{\smallskip}
\noalign{\smallskip}
Observation & Telescope     & {\it p} & $\mu$ & $\sigma$ & $\chi^{2}$ & $N_{d.o.f}$ & $P(\chi^{2})$ \\
day         & \& component  &         &       &          &            &             &               \\
\noalign{\smallskip}
\hline
\noalign{\smallskip}
53886/31.05.06 & Lovell M1     &       &  2.28 & 0.26 &   346 & 125 & $3 \times 10^{-4}$\\
               & Lovell M2     & -2.81 &       &      &   463 & 176 & $7 \times 10^{-5}$\\
               & WSRT M1       &       & 11.10 & 0.63 &   304 &  95 & $3 \times 10^{-5}$\\
               & WSRT M2       & -2.35 &       &      &   130 &  62 & $2 \times 10^{-5}$\\
               & WSRT M3       & -2.15 &       &      &   191 &  86 & $9 \times 10^{-3}$\\
               & Effelsberg M1 & -1.85 &       &      &   390 & 149 & $3 \times 10^{-8}$\\
               & Effelsberg M2 & -1.89 &       &      &   438 & 125 & $2 \times 10^{-5}$\\
               & Effelsberg M3 & -1.41 &       &      &   448 & 147 & $4 \times 10^{-10}$\\
\noalign{\smallskip}
\hline
\noalign{\smallskip}
53926/10.07.06 & Lovell M1     &       &  1.91 & 2.99 &   382 & 107 & $9 \times 10^{-3}$\\
               & Lovell M2     & -7.68 &       &      &   285 & 103 & $8 \times 10^{-10}$\\
               & Eff M1 Strong &       &  5.38 & 2.88 &   120 & 123 & $3 \times 10^{-3}$\\
               & Eff M1 Weak   &       & -0.49 & 3.27 &  1704 & 121 & $8 \times 10^{-9}$\\
               & Effelsberg M2 &       & -0.54 & 0.53 &   357 &  90 & $7 \times 10^{-2}$\\
               & Eff M3 Strong & -0.89 &       &      &    41 & 123 & $3 \times 10^{-1}$\\
               & Eff M3 Weak   &       & -7.29 & 1.25 &   402 & 103 & $7 \times 10^{-3}$\\
               & Effelsberg IP &       & -1.97 & 2.36 &  1446 &  93 & 0.0\\
\noalign{\smallskip}
\hline
\noalign{\smallskip}
53933/17.07.06 & Lovell M1     &       &  1.46 & 0.97 &   423 & 116 & $3 \times 10^{-11}$\\
               & Lovell M2     & -2.10 &       &      &   558 &  92 & $1 \times 10^{-10}$\\
               & WSRT M1       & -2.86 &       &      &   279 &  74 & $2 \times 10^{-7}$\\
               & WSRT M2       & -2.86 &       &      &   281 &  84 & $5 \times 10^{-12}$\\
               & WSRT IP       & -1.78 &       &      &   440 & 261 & $6 \times 10^{-6}$\\
               & Eff M1 Strong$^{\ast}$ & $-$   & $-$   & $-$    & $-$  & $-$ & $-$ \\
               & Eff M1 Weak   &       & -1.98 & 0.99 &   463 & 103 & $1 \times 10^{-12}$\\
               & Effelsberg M2 &       & -0.59 & 0.48 &   276 &  98 & $1 \times 10^{-3}$\\
               & Eff M3 Strong$^{\ast}$ & $-$   & $-$   & $-$    & $-$  & $-$  & $-$ \\
               & Eff M3 Weak   &       & -0.11 & 4.12 &   557 &  83 & 0.0 \\
               & Effelsberg IP &       & -1.86 & 2.41 &  1241 & 118 & 0.0 \\
\noalign{\smallskip}
\hline
\end{tabular}
\smallskip
\begin{list}{}{}
\item[$^{\ast}$] insufficient number of pulses to do the fitting
\end{list}
\end{minipage}
\end{table*}

Session 3 is the only one, where the IP becomes present in all the data sets. Unfortunately, the IP subpulses from the 1.4 GHz Lovell data set are not seen above the noise level in the pulse-energy distribution, despite their appearance in the pulse stack. This is due to the weak S/N ratio and interference present in that data set. The M1 from the Lovell is best fit by a lognormal distribution, while M2 is best fit by a power-law distribution. The 4.9 GHz WSRT pulse-energy distributions show a power-law-like behaviour in all the components. A very strong IP is visible in the 8.35 GHz Effelsberg data set, its pulse energy distribution is similar to the data set from session 2. The MP and the IP pulse-energy distributions show lognormal-like behaviour. Similarly to the previous session both the M2 and M3 pulses show a complex pulse energy distribution and the IP shows a peak, indicating many pulses with flux close to the average energy of the IP, and a long tail in the pulse-energy distribution. Especially the Effelsberg data set shows long tails. Considering only the Effelsberg strong pulses above 5 times the average energy of the component, their pulse-energy distribution is statistically insignificant because of the low number of pulses. The fit to the IP data from Effelsberg is also not significant.

\begin{figure*}
\includegraphics[scale=1.0,angle=0]{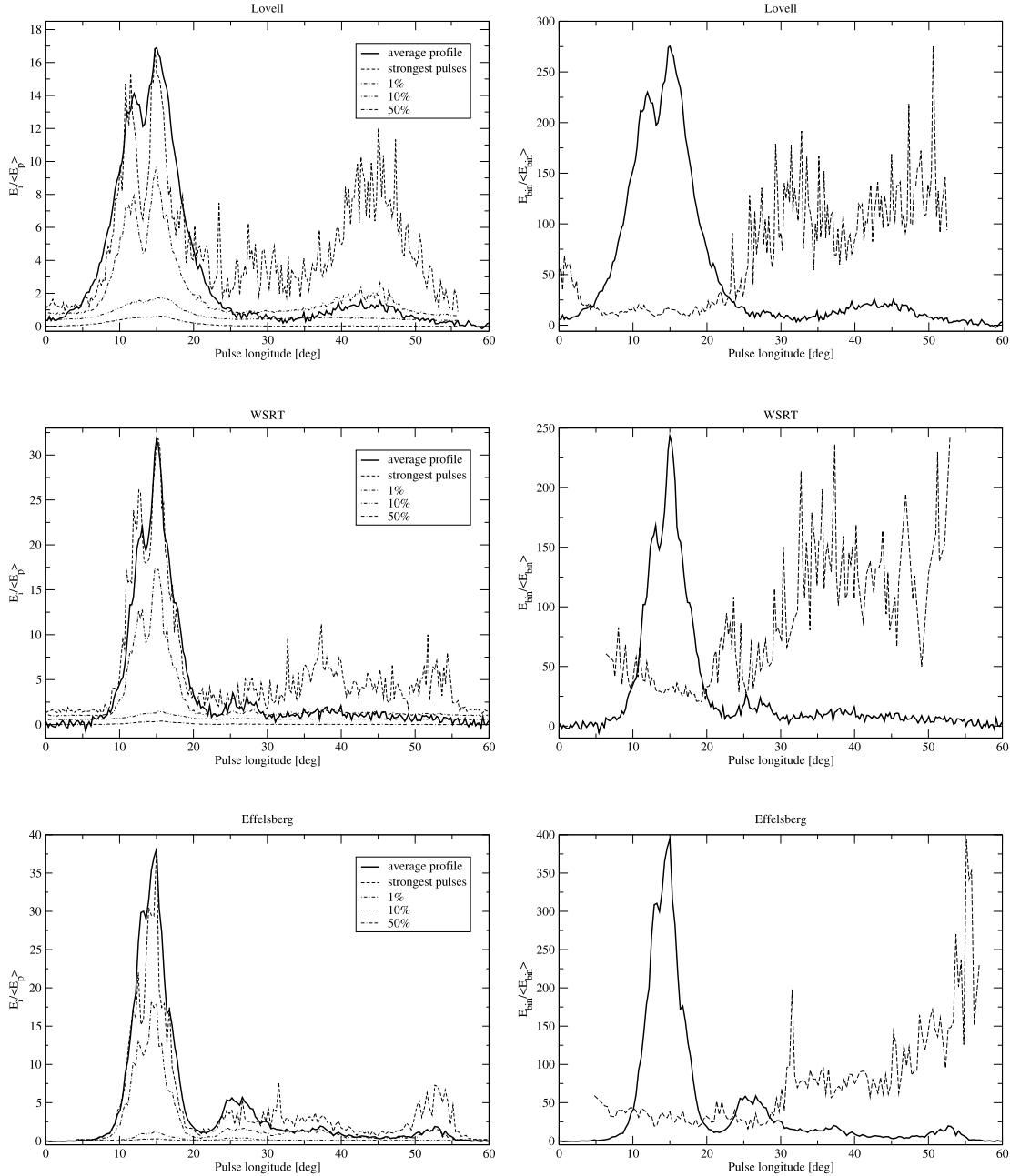}
\caption{The contour plot of the longitude-resolved cumulative pulse-energy distributions from the 1.4 GHz Lovell ({\it top row}), 4.9 GHz WSRT ({\it middle row}) and 8.35 GHz Effelsberg ({\it bottom row}) radio telescope observations from the session 1. {\it Left:} the thick solid line is the average pulse profile. The dashed line shows the brightest time sample for the each pulse-longitude bin. The other lines are the contours of the longitude-resolved cumulative pulse-energy distributions at the energy level of the 1\%, 10\% and 50\% compared to the average peak energy $<E_{p}>$ of the average pulse profile. {\it Right:} the thick solid line is pulse profile. The dashed line shows the brightest time sample for each pulse-longitude bin compared with the average intensity $<E_{bin}>$ at that pulse longitude.}
\label{fig:all_telescope_contour_53886}
\end{figure*}

To better illustrate the strong intensity fluctuations of XTE J1810--197 and to investigate any longitude-dependence of its strong subpulses, we have produced contour plots of the longitude-resolved cumulative pulse-energy distributions. Fig.~\ref{fig:all_telescope_contour_53886} (left column) presents the distributions calculated for all the telescopes from session 1. The average profile is drawn with the thick line, while different linestyles denote the contours of the cumulative pulse-energy distribution at the energy levels of 1\%, 10\% and 50\%, normalised by the average peak energy $<E_{p}>$ of the average pulse profile. The 10\% contour, for example, shows the energy level of the pulse-energy distribution above which 10\% of pulses can be found. The second type of plot (right column; Fig.~\ref{fig:all_telescope_contour_53886}) indicates the brightest time samples of each pulse-longitude bin (dashed line) scaled to the average intensity at that pulse-longitude bin, plotted on top of the average pulse profile (solid line).

In the 1.4 GHz Lovell observation of session 1, subpulses appearing in the first component do not vary much, which indicates stable emission, this behaviour changes in M2 (Fig.~\ref{fig:all_telescope_contour_53886}; left column, top row) and this behaviour is repeated in session 2 where it becomes very weak, although there are still some spiky subpulses present. The 4.9 GHz WSRT observation shows strong subpulses in M1 only, and the emission is stable throughout the rest of the profile during all sessions.

\begin{figure*}
\includegraphics[scale=1.0,angle=0]{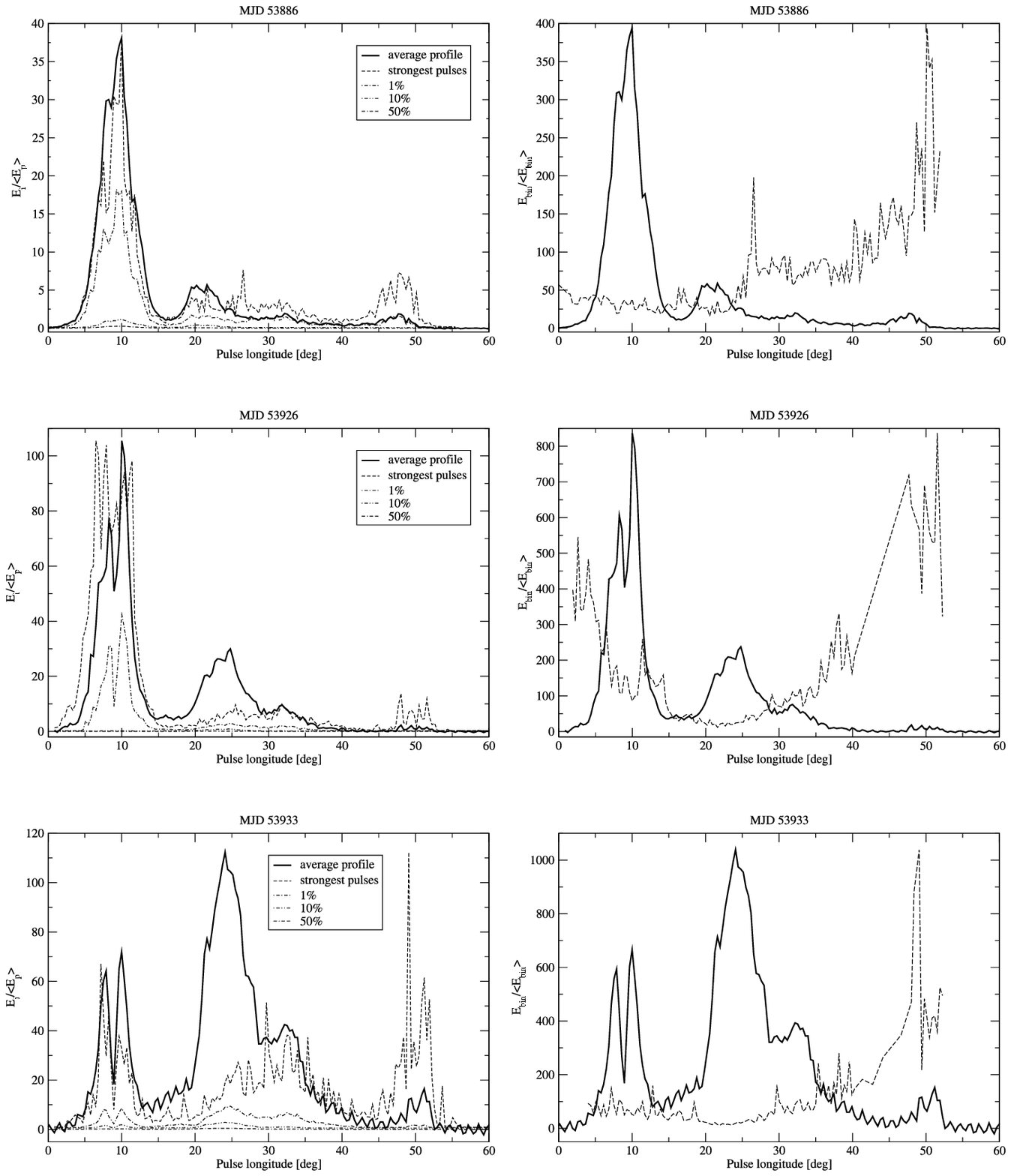}
\caption{The contour plot of the longitude-resolved cumulative main pulse energy distributions from the 8.35 GHz Effelsberg radio telescope observations from session 1 ({\it top row}), session 2 ({\it middle row}) and session 3 ({\it bottom row}). The notation is the same as in Fig.~\ref{fig:all_telescope_contour_53886}.}
\label{fig:effberg_contour_all_days_mp}
\end{figure*}

To illustrate the session-to-session changes we used observations from the Effelsberg telescope. The strong signal and the presence of the IP, as can be seen in Fig.~\ref{fig:effberg_contour_all_days_mp}, show how the emission from different components changes in time. M1 in the MP is always strong and the contour plots show a strong stable emission (shown by 50\% energy level contours). It is also worth noticing that there are two peaks emerging in time. M2 grows stronger from session to session, and it always shows stable and strong emission. M3 is weak and is dominated by the presence of a few strong spiky subpulses.

\subsection{Single pulse correlation}

The last stage of our analysis was to perform a single pulse correlation. Before the analysis, we needed to align the data. The reason for this is the interstellar dispersion which causes the pulses received at higher frequencies to arrive earlier than the lower frequencies and the different path-lengths to the telescopes. Therefore all pulse arrival times have been corrected to a common reference frame. We transfered the times of the pulses to the solar system barycentre, using the DE200 ephemeris \citep{sta82}. Then, the overlapping data sets at the different frequency pairs were taken for alignment, which was done by comparing the times of arrival of the pulses. Each data set was aligned in time with a one phase bin accuracy for each frequency pair. With the time resolution of 5.4 ms, the data sets were sufficiently aligned and all the telescope-specific delays were negligible. We applied two techniques of single pulse correlation, the longitude-resolved linear correlation (LRLC) and the longitude-resolved cross-correlation (LRCC). Both methods are complementary and give interesting results as we will show below.

The longitude-resolved linear correlation is based on a method first introduced by \citet{pop86}. In order to produce the so-called 'linear correlation map' of the pulse-to-pulse variations, the linear correlation array $C_{i,j}$ as presented in the work of \citet{khk+01} needs to be calculated according to the following formula:

\begin{equation}
C_{i,j} = \frac{1}{n \times \sigma_{f,i} \times \sigma_{g,j}}\displaystyle\sum_{k=1}^{n}{[f_{i}(k) \times g_{j}(k)-<f_{i}><g_{j}>]},
\label{eq:linear correlation}
\end{equation}

\noindent where $f_{i}(k)$ and $g_{i}(k)$ are time series of a bin {\it i} from two distinctive frequencies with {\it k} being the pulse number, while $\sigma_{f,i}$ $\sigma_{g,j}$ being their standard deviations respectively.

\begin{figure*}
\includegraphics[scale=0.8]{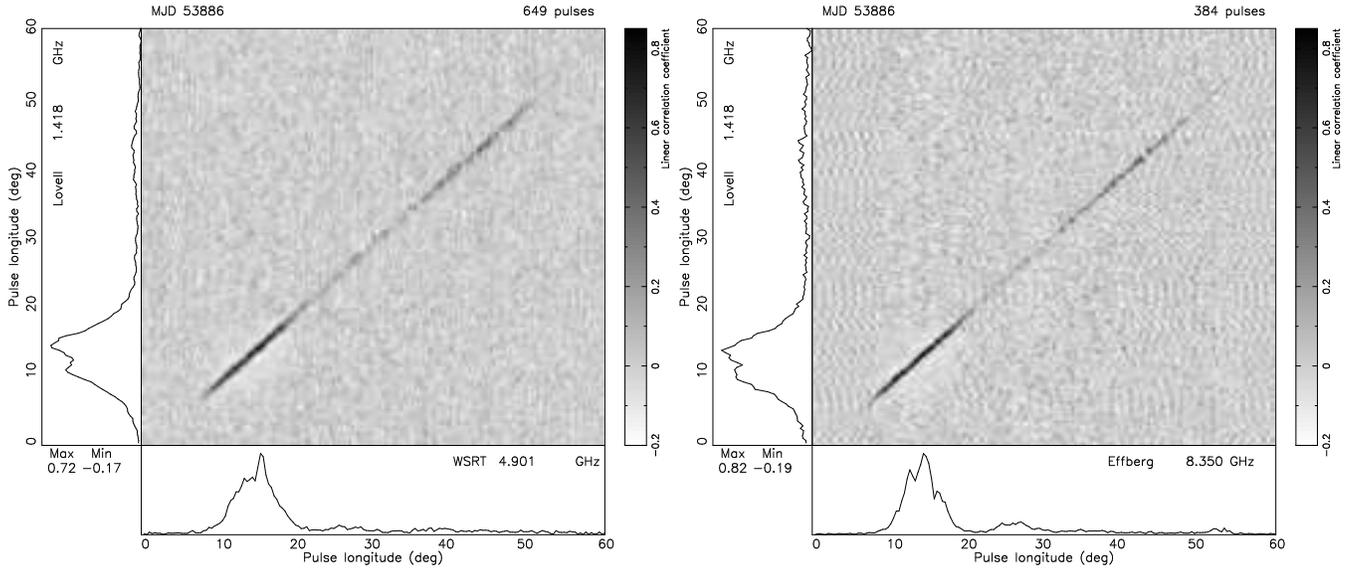}
\caption{The plots of linear correlation between the 1.4 GHz Lovell and 4.9 GHz WSRT ({\it left}), and the 4.9 GHz WSRT and 8.35 GHz Effelsberg data sets ({\it right}) from session 1. The contrast of the image was increased by showing only part of the cross-correlation coefficient range. The bottom and left panel hold the average profiles of the correlated data sets. The bar on the right side of the plot denotes the different levels of the linear correlation coefficient. The value of -0.2 means 20\% negative linear correlation while 0.85\% positive linear correlation is denoted by 0.85 value. The maximal and minimal values of the linear correlation coefficients are shown in the left lower part of plots. Characteristic wave-like pattern visible in the lower plot is a result of the correlation of the signal from the Lovell with the RFI present in the Effelsberg data set.}
\label{fig:linear correlation maps}
\end{figure*}

The region of the highest linear correlation is expected to fall on the region around a diagonal which represents the time and spatial scale along which the subpulse intensities are linearly correlated. The LRLC method provides a good insight into linear dependent intensity variations of individual pulses at different observed frequencies. However it does not take into account phase, shape information and non-linear dependencies between the components of the pulse from the different frequencies. The presence of low-intensity but persistent RFI in one of the data sets causes the appearance of a wave-like pattern in the linear correlation map as can be seen in the lower panel of Fig.~\ref{fig:linear correlation maps}.

An example of the total intensity linear correlation between the 1.4 GHz Lovell and 4.9 GHz WSRT (upper plot) and the 1.4 GHz Lovell and 8.35 GHz Effelsberg (lower plot) data sets can be seen in Fig.~\ref{fig:linear correlation maps}. The linear correlation of M1 is always strong between all frequency pairs. Most of the correlation regions in our results are very narrow. This confirms the results from previous analysis steps, of strong and moderately stable emission from M1 at all frequencies and at all epochs. Surprisingly, the linear correlation of M2 is not as high despite the results showing it as the region with lowest intensity fluctuations which is characterised by low modulation indices in all our analysis. It is also interesting to note that the correlation seems to increase again towards the trailing edge of the pulse profile. The IP is visible only in the last two sessions, however in the case of session 2 the IP appears only in the 8.35 GHz Effelsberg data set. This constrains the IP correlation analysis to the last session only. Unfortunately, the IP in the 1.4 GHz Lovell data set is very weak and the correlation results comprising this data set are unreliable. The linear correlation of the remaining data sets is equally strong across the whole IP range and higher than that of MP.

\begin{figure*}
\includegraphics[scale=0.8]{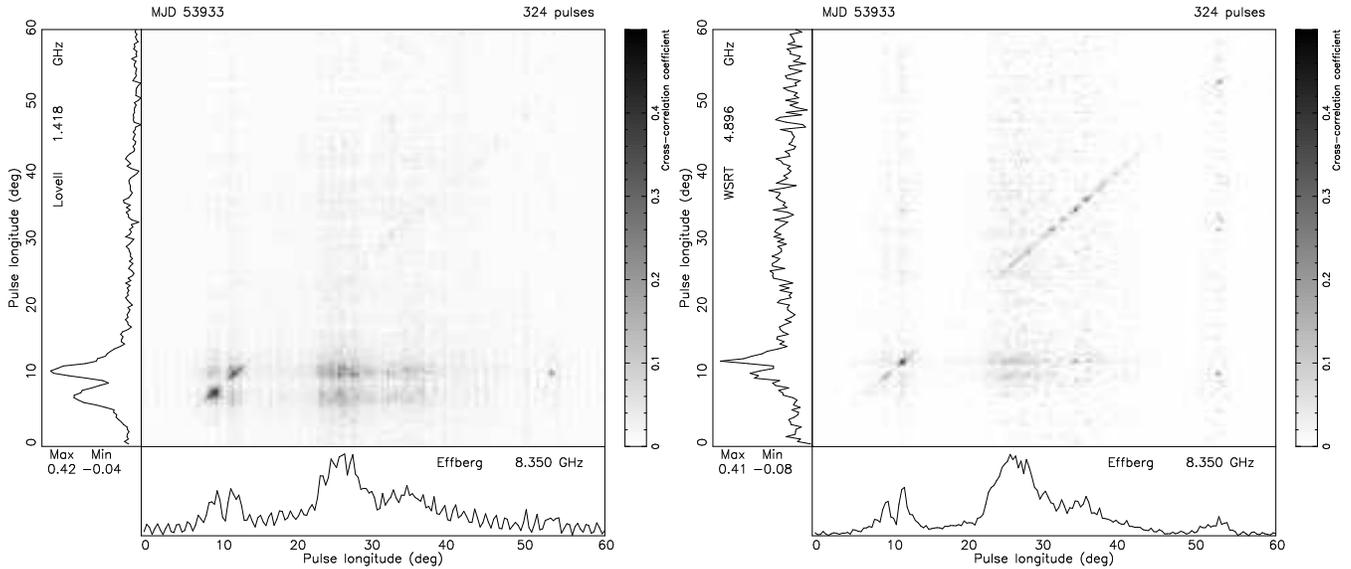}
\caption{The plots of cross-correlation between the 1.4 GHz Lovell and 8.35 GHz Effelsberg ({\it left}), and the 4.9 GHz WSRT and 8.35 GHz Effelsberg data sets ({\it right}) from session 3. The contrast of the image was increased by showing only part of the cross-correlation coefficient range. The profiles in the bottom and left panel represent the average profiles. The bar on the right side of the plot denotes the different levels of the cross-correlation coefficient normalised for clarity. The value of 0.0 means no cross-correlation while 50\% positive cross-correlation is denoted by 0.5 value. The maximal and minimal values of the cross-correlation coefficients are shown in the left lower part of the plots.}
\label{fig:cross-correlation maps}
\end{figure*}

Information on the shape and non-linear dependencies can be obtained by applying the LRCC method. This method of analysis is also more resistive to the presence of the periodic RFI in the 8.35 GHz Effelsberg data. The principle of producing the cross-correlation map is somewhat similar to the LRLC but instead of Eq.~\ref{eq:linear correlation} we first apply the Fast Fourier Transform to both data sets. The next step is to multiply the first transformed data set by the complex conjugate of the second one. This produces a complex result of which the imaginary part is zero. Taking only the real part of the resulting cross-correlation coefficients matrix, the cross-correlation map is made as in the LRLC method. Examples of the results from the LRCC are presented in Fig.~\ref{fig:cross-correlation maps}. The results of the cross-correlation analysis are similar to that of linear correlation. In session 3 we noticed the presence of a weak cross-correlation between the different components. Further analysis revealed that the occurrence of regions of higher correlation outside diagonal is caused by the RFI present in the 8.35 GHz Effelsberg data set. The region of correlation along the diagonal is very narrow in all of the correlated data sets. Similarly to the LRLC the cross-correlation is strong in M1 in all of the LRCC results, and can be explained by the similarities in shape and of the components at different frequencies. The next two components M2 and M3, are well correlated only between 4.9 and 8.35 GHz data sets (Fig.~\ref{fig:cross-correlation maps}, right panel). The correlation results of 1.4 GHz with 4.9 or 8.35 GHz data sets do not show any correlation in M2 and M3 components.

\section{Summary and conclusions}

We have analysed the data from the simultaneous multi-frequency single-pulse observations of XTE J1810--197 conducted at the frequencies of 1.4, 4.9 and 8.35 GHz, during observing sessions in May and July 2006. The phenomena revealed by our analysis indicate that the radio emission is clearly different to the known radio pulsar properties. Previous work by \citet{ccr+07}, \citet{crp+07}, \citet{ksj+07} and \citet{ljk+08} has shown an interesting overview of peculiarities of the pulsed radio emission from the magnetar, e. g. flat spectral index, $\alpha = 0.0 \pm 0.5$, high degree of polarisation or long-term evolution of the polarisation angle swing with time. The results presented in our work confirm that the mechanism responsible for the magnetar radio emission appears to have a different origin or perhaps even multiple origins, compared to the normal radio pulsars.

XTE J1810--197 has a broad, multi-component profile with the IP only becoming present after session 1. Results from \citet{ksj+07} show that the separation between the MP and the IP is less than 180 deg which may be caused by an extremely wide MP beam ($\rho \sim 44$ deg) or by a non-dipolar magnetic field structure. The evolution of the average pulse profile takes place on day-to-day time-scales. This is clearly visible in the highest frequency data where the first component decreases in intensity, while the second and third become more prominent. \citet{ksj+07} postulate, after examining the average pulse profiles of XTE J1810--197 from a greater number of observing sessions, that the magnetar requires an unusually long observing time to obtain a stable profile or even lacks one, perhaps like PSR B0656$+$14 known for its bursting behaviour \citep{wws+06}.

The modulation index values change dramatically from one component to another even within a single observing frequency. \citet{wse07} define the modulation index to be the longitude-resolved modulation index $m_{i}$ at the pulse longitude bin {\it i} where $m_{i}$ has its minimum value. In most of the sources used in their analysis the LRMI shows a minimum in the middle of the pulse profile with a typical value of $m \sim 0.5$, where the total intensity is relatively high. We also find that tendency in our results, but the magnitude of the values is somewhat closer to the values reported for the Crab Pulsar ($m = 5$). This agrees with the conclusion that due to the infrequent occurrence of the strong subpulses with narrow widths and broad distribution of the subpulse intensities, the modulation index is on average significantly larger in XTE J1810--197. We must note however, that in few cases, frequent occurrence of many strong and narrow subpulses with similar intensities at certain pulse phases of MP, results in lower LRMI values. The lowest values of the modulation indices occur during session 1, but even in this session they increase with increasing frequency as can be seen in Fig.~\ref{fig:all_telescope_2dfs_53886} (upper panels). Such behaviour is in contrast to the normal radio pulsars where the LRMI values from lower frequencies are on average higher than that at higher frequencies \citep{wse07}. As we move in to session 2 the intensity fluctuations grow stronger with minimum modulation indices at values of around 4 for the 1.4 GHz Lovell data sets. For the 8.35 GHz Effelsberg data, we calculate the minimum modulation indices to be 1 in the MP and 2 in the IP region. It is remarkable, that for both data sets in this session the modulation indices in the MP peak at values of 7.5 and 10 for the Lovell and Effelsberg data respectively. These values are extremely high and except for the Crab Pulsar, unprecedented in the results from modulation analysis from normal radio pulsars \citep{wse07}. In session 3, for the 4.9 GHz WSRT data sets due to the low signal-to-noise ratio, the LRMI values are not sampled densly throughout the whole pulse profile range, but available values are comparable with session 1. The Lovell and Effelsberg LRMI values are similar to that of session 1. The longitude-resolved modulation analysis results presented above show the variation of the LRMI values on day-to-day time-scales and dramatic changes with pulse phase. In all session we find the frequency evolution of the LRMI values in contradiction with proprieties of normal radio pulsars.

The following step of our analysis, the 2DFS, do not show any regular drifting behaviour in any sessions or any frequencies. However, we find phenomenon manifesting itself as characteristic bumps in the XTE J1810--197 vertically collapsed 2DFS. We interpret this phenomenon as the tendency of the subpulses to be equally separated in the consecutive pulses' profiles throughout observation. We also notice the presence of a peak visible in the vertically collapsed 2DFS at 1.4 GHz Lovell, 4.9 GHz WSRT from session 1 (Fig.~\ref{fig:all_telescope_2dfs_53886}, left and middle panels) and 8.35 GHz Effelsberg from session 3 (Fig.~\ref{fig:effberg_2dfs_all_days}, right panel in the upper row) data sets. We interpret the peak as a signature of the baseline variations in those data sets. The lack of regular drift from the magnetar might be associated with its rapidly changing emission properties and young age ($\tau < 10$\,kyr). In their work, \citet{wse07} have shown that the fraction of young pulsars showing regular drifting is very low. Although, one could also argue that the strong radio variability might mask any regular structures or the physics of the magnetar's radio emission is different from radio pulsars.

The high variability of the magnetar emission is also reflected in its pulse-energy distributions. For all sessions at all observed frequencies we have made pulse-energy distributions for each of the main-pulse components as well as for the IP (whenever present). We fit each of the pulse energy distributions with models based on a power-law or lognormal statistics for comparison between our observations and existing pulsar emission models. As justified later in this section, propagation effects in the interstellar medium are negligible for our data analysis and are therefore ignored. Table~\ref{table:pulse energy fits} presents the results of the best fits. The significance of the best fits are low due to the oversimplified models, but changes in the best-fit models of the components in different sessions are nevertheless, very peculiar. We interpret that as indicating the possible presence of multiple emission mechanisms with different statistical behaviour embedded in the same pulse phases.

In a series of papers, Cairns and collaborators (\citealt{cjd01}, \citealt{cjd03a}, \citealt{cdr+03}, \citealt{cjd04}) investigate different possible models of the emission physics using observations of PSRs B0833--45 (Vela Pulsar), B1641--45 and B0950$+$08. They show that the models of emission vary with changing pulse phase in the analysed sources. They find in these pulsars, that longitude-resolved pulse-energy distributions which appear to be lognormal and are representative of the normal pulsar radio emission, can be fit with a single emission model, or convolution of Gaussian-lognormal or double lognormal models. This multi-model manifestation is explained as two non-related waves coupling together in the inhomogeneous plasma in the pulsar magnetosphere. While Carins et al. successfully fit the longitude-resolved pulse-energy distributions with lognormal statistics, in many cases it is also valid to use it to fit the integrated pulse-energy distributions \citep{jr02}. The presence of approximately power-law distributions is caused by these different phenomena. The origin of these phenomena can be identified based on the values of power-law indices and can be associated with (i) giant pulses i. e. pulses which have integrated flux density greater than 10 times the integrated flux density of the average profile, (ii) giant micropulses, which have a peak flux density that is 10 times the peak flux density of the average profile at the same phase, but less than 10 times integrated flux density of the average profile, (iii) so-called precursor emissions occurring just before the MP as for Crab Pulsar (\citealt{cbh+04}, \citealt{mh96}), which have power-law distributions that are non intrinsic and caused by Gaussian distributions from the background normal pulsar emission convolved with higher energy lognormal components. We argue that in the case of the magnetar, M1 consists of strong and narrow giant-like subpulses followed by a components of weak precursor-like subpulses.

The giant pulse and giant micropulse phenomena are known in a few radio pulsars like PSRs B0531$+$21 (Crab Pulsar; \citealt{sr68}), B0833--45 (Vela Pulsar; \citealt{jvkb01}), B1937$+$21 \citep{spb+04} or B1133$+$16 \citep{kkg+03}. The emission of these pulses can be characterised by the power-law energy distributions, broadband emission or occurrence within narrow pulse phases. These phenomena are believed to be associated with the high-energy emission in the outer magnetosphere (\citealt{jr02}, \citealt{cai04}). In the case of XTE J1810--197 we also see strong spiky subpulses which could be associated with the giant pulse phenomenon, but their widths are larger than that of giant pulses of normal radio pulsars. Also, their occurrence, which covers the whole longitude range of that component, stand in contradiction to this definition. The most prominent example illustrating the above phenomenon in our observations is the first component, especially in the Effelsberg data sets from sessions 2 and 3 as can be seen in Fig.~\ref{fig:effberg_penergy_all_days_mp}. This component has many strong and spiky subpulses appearing within its whole longitude range, which dominates the high-energy tail in its pulse-energy distributions. This makes fitting the pulse-energy distributions with only one law impossible. However, the attempt to decouple the component distribution into low and high energy parts also did not result in good fits. A similar case occurs for PSR B0656$+$14 \citep{wse07}, where weak emission is coupled with a component responsible for high energy bursts. In the case of XTE J1810--197 there may be more than two models contributing at one phase in the observed distributions. As shown later the magnetars' very dynamic magnetosphere may be an explanation for such multi-component pulse-energy distributions. This argues that the emission is in general broadband, but the degree of variability is very different. Those components average together in pulse-energy distributions, which makes them difficult to fit properly using known statistical models.

Despite the changes in the pulse profile and pulse-amplitude characteristics on short time-scales, the correlation analysis gave results which contradict the overall picture of unstable emission from XTE J1810--197. The LRLC analysis shows significant correlation results in the majority of the frequency pairs used. The narrow and high correlation regions denote significant dependence between the intensities of the subpulses on small time- and spatial scales. The correlation always occurs between the first components in all of the analysed frequency pairs, with sporadic correlation between the third components. In contrast, the second component, is found to be a stable emission region with lower modulation indices, was very weakly correlated. To examine the non-linear dependence between frequencies we used the LRCC method. The correlation is weaker when compared to the LRLC method, but correlated regions are also very narrow, showing that there are similarities in the phase and shape of the subpulses at different observed frequencies. 

While the larger time-scale flux variations might be normally attributed to interstellar scintillation, they have been rejected in the work of \citet{crp+07} and \citet{ljk+08} as being responsible for the variability over short time scales. This points to the intrinsic behaviour of the magnetar as a cause. The lack of a regular drift, broad pulses, the presence of subpulses with quasi-periodic modulation, difficulties with fitting the data with single lognormal or power-law models allow us to draw a conclusion of non-stable emission due to the possible turbulent magnetar magnetosphere. A model explaining that emission has been proposed very recently. In his work, Thompson (\citealt{tho08a,tho08b}) gives an extensive explanation of the pair creation processes in ultra-strong magnetic field and particle heating in a dynamic magnetosphere. He considers the details of the QED processes that create electron-positron pairs in high magnetic fields of the order of $10^{14}$\,G. He discusses the possibility of a strong enhancement of the pair creation rate in the open-field circuit and outer magnetosphere by instabilities near the light cylinder. Thompson also refers to the flat radio spectra as a possible result of the high plasma density in the open magnetic field lines. One of the model explanations of the magnetar's broad pulse profile, is its beam geometry. In normal radio pulsars, wide pulse profiles are usually caused by the alignment between rotation and magnetic pole axis. The line of sight of the observer stays within the emission beam for a large fraction of the pulse period resulting in the long duty cycle. In the case of XTE J1810--197 the solution of fitting the position angle swing with the Rotating Vector Model \citep{rc69a} results in non-aligned geometry ($\alpha = 44$ deg,\,$\beta = 39$ deg), but the beam radius inferred from the MP pulse has a width of about $\rho \sim 44$ deg as shown by \citet{ksj+07}. This result excludes viewing geometry as a reason for a wide pulse profile in XTE J1810--197. The model of the dynamic outer magnetosphere has a promising application in explaining the radio emission from the magnetars and is consistent with the magnetars' emission features such as flat radio spectra, broad pulses and rapid variability.

Since the detection of radio emission from XTE J1810--197, its relation to the new class of objects called Rotating RAdio Transients (RRATs, \citealt{mll+06}) was suggested. The lack of X-ray counterpart in any of known RRAT sources argued against this hypothesis until \citet{rbg+06} reported the first X-ray detection from RRAT J1819-1458. This RRAT was known to emit radio bursts with 4.26\,s spin period based on its timing analysis and the inferred dipole surface magnetic field of about $5\times10^{13}$\,G. The detection of the periodic X-ray pulsations aligned with the radio bursts by \citet{mrg+07} allowed the comparison of its X-ray emission proprieties with XTE J1810--197 and excluded a close relationship with this RRAT. However, it would be very interesting as proposed by \citet{rmg+08} to search for the RRAT-like radio bursts from XTE J1810--197 at its quiescence level. This could investigate the hypothesis of a link between the magnetars, RRATs and young radio pulsars.

\section*{Acknowledgements}
We are very grateful to Ramesh Karuppusamy for valuable suggestions and fruitful discussions during our data analysis. We also thank Gemma Janssen for her help with our WSRT observations. Maciej Serylak was supported by the EU Framework 6 Marie Curie Early Stage Training programme under contract number MEST-CT-2005-19669 "ESTRELA". Kosmas Lazaridis was supported through a stipend from the International Max Planck Research School (IMPRS) for Astronomy and Astrophysics at the Universities of Bonn and Cologne.

\label{lastpage}
\end{document}